\documentclass{pasj00}

\usepackage{times}

\newcommand{\cleft}[1]{
  \multicolumn{1}{l}{#1} 
}

\begin{document}
\SetRunningHead{Yamamoto et al.}{Cyclotron Resonance Feature in GRO J1008$-$57}

\Received{2013/11/29}
\Accepted{2014/01/18}

\title{ Firm Detection of a Cyclotron Resonance Feature  with Suzaku \\
 in the X-ray  Spectrum of 
  GRO J1008$-$57 during a Giant Outburst in 2012}

\author{%
  Takayuki \textsc{Yamamoto},\altaffilmark{1}
  Tatehiro \textsc{Mihara},\altaffilmark{1}
  Mutsumi \textsc{Sugizaki},\altaffilmark{1}
  Motoki \textsc{Nakajima},\altaffilmark{2}
  Kazuo \textsc{Makishima},\altaffilmark{1,3}\\
  and
  Makoto \textsc{Sasano}\altaffilmark{3}
}

\altaffiltext{1}{MAXI team, RIKEN, 2-1 Hirosawa, Wako, Saitama 351-0198}
\email{tyamamot@crab.riken.jp}
\altaffiltext{2}{School of Dentistry at Matsudo, Nihon University, 
  2-870-1 Sakaecho-nishi, Matsudo, Chiba 101-8308}
\altaffiltext{3}{Department of Physics, The University of Tokyo, 
  7-3-1 Hongo, Bunkyo, Tokyo 113-0033}


\maketitle

\begin{abstract}

  We report on the firm detection of a cyclotron resonance scattering
  feature (CRSF) in the X-ray spectrum of the Be X-ray binary pulsar,
  GRO J1008$-$57, achieved by the Suzaku Hard X-ray Detector
  during a giant outburst which was detected by the MAXI Gas Slit Camera
  in 2012 November.  The Suzaku observation was carried out on 2012 November 20, 
  outburst maximum when the X-ray flux reached $\sim 0.45$ Crab
  in 4--10 keV, which
  corresponds to a luminosity of $1.1\times 10^{38}$ erg s$^{-1}$ in 0.5--100 keV at 5.8 kpc.
  The obtained broadband X-ray spectrum from 0.5 keV to 118 keV
  revealed a significant absorption feature, considered as the
  fundamental CRSF, at $\sim 76$ keV.  This unambiguously reconfirm the previously suggested 
  $\sim$ 80 keV spectral feature in GRO J1008$-$57.
  The implied surface magnetic field,
  $6.6\times 10^{12}$ G, is the highest among binary
  X-ray pulsars from which CRSFs have ever been detected.
  
\end{abstract}

\section{Introduction}
\label{sec:intro}

%
Pulsars, which exhibit pulsating electromagnetic radiations
in various wavelength, are strongly magnetized neutron stars.
The rotation of the neutron star, combined with anisotropic radiation,
causes the periodic pulsation.  
Although they are considered to be
formed by supernova explosions of massive stars, the origin and time evolution of their
magnetic fields are still open questions.

X-ray binary pulsars (XBPs) are a group of X-ray binaries involving
pulsating neutron stars.  According to the type of the binary
companion, they are classified into several subgroups including Super
Giant XBPs and Be XBPs as major members (e.g. \cite{Reig2011}).  
Be XBPs produce recurrent outbursts synchronized with their binary orbital periods.
The outbursts are considered to occur when the neutron star crosses a
gaseous stellar disk of the Be star near the periastron passage. The outburst
does not always appear every orbital cycle, and sometime arises in an
irregular orbital phase, probably depending on the physical extent of
the stellar disk.
%

Surface magnetic fields of neutron stars in XBPs can be estimated from
the cyclotron resonance scattering feature (CRSF), which has been
observed as absorption features in their X-ray spectra.  
The CRSF is considered to appear at an energy of 
$E_{\rm a} = 11.6 (1+z_{\rm g})^{-1} B_{12}$,
where 
$B_{12}$ is the magnetic field strength in $10^{12}$ G, and 
$z_{\rm g}$ represents the gravitational redshift.
Ginga/LAC observations in the 2--60 keV band detected the CRSFs from 12
XBPs and showed that their surface magnetic fields are distributed in a very
narrow range of $(1.0-3.2)\times 10^{12}$ G (\cite{Mihara1998}, \cite{Makishima1999}).
%
Subsequently, ASCA, RXTE, BeppoSAX, INTEGRAL and Suzaku
observations surveyed a wider energy band from $\sim 0.5$ keV upto a
few hundreds keV, and detected CRSFs from additional six XBPs
(e.g. \cite{Coburn2002}; \cite{Filippova2007}; \cite{Doroshenko2010};
\cite{Yamamoto2011}; \cite{Tsygankov2012}; \cite{DeCesar2013}).
However, the revised range of their surface magnetic fileds,
$(1.0-4.7)\times 10^{12}$ G, is still narrow.
It is yet to be clarified 
whether this is intrinsic to XBPs, or a selection effect due to limited observations.

GRO J1008$-$57 is a Be XBP with a pulsation period of 93.5 s,
discovered by the CGRO/BATSE in 1993 (\cite{Stollberg1993}).  Its optical
counterpart was identified with a B0e type star \citep{Coe1994} and the
distance was estimated to be 5.8 kpc \citep{Riquelme2012}.  Its X-ray
outbursts have been monitored for about 20 years by surveys with the CGRO/BATSE, RXTE/ASM, Swift/BAT, and
MAXI/GSC.  Since 2003 January, the source
has been in an active state exhibiting outbursts periodically
\citep{Kuhnel2013}.  From the recurrent outburst intervals and the
pulsar period modulation, the binary orbital period was estimated 
as $247.8\pm 0.4$ d \citep{Coe2007}, which was recently refined to
249.48 $\pm$ 0.04 d by the pulse arrival-time analysis
\citep{Kuhnel2013}.
Based on 
the CGRO/OSSE pointing observations performed in the 1993 outburst and the BATSE
earth-occultation data on that occasion, 
\citet{Shrader1999} suggested a possible CRSF at around 88 keV
in the X-ray spectra of GRO J1008$-$57.
In contrast, spectra of the 2004 outburst obtained by
the INTEGRAL/IBIS and JEM-X showed no feature in the 3--60 keV band
\citep{Coe2007}.  Therefore, the possible absorption feature at 88 keV
is considered to be the fundamental if it is real.  Observations of
the 2007 November outburst by RXTE, Swift, and Suzaku 
were unable to confirm the suggestion, hampered by
rather poor signal statistics \citep{Naik2011,Kuhnel2013}.


In the present paper, we report the Suzaku observation performed at
the peak of a giant outburst detected by MAXI in 2012
November, and the results of the spectral analysis for the CRSF.
Unless otherwise specified, all errors hereafter refer to 90\% confidence limits.



\section{Outburst Activity Monitored by MAXI}
\begin{figure}
  \begin{center}
    \includegraphics[width=85mm]{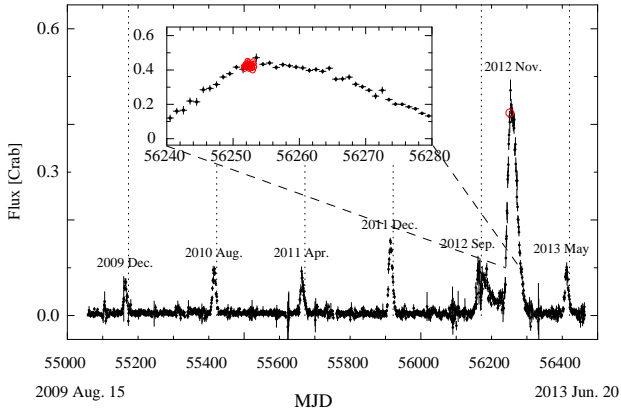}
  \end{center}
  \caption{ MAXI/GSC light curve of GRO J1008$-$57 in the 4--10 keV band
    from 2009 August 15 to 2012 June 20.  Vertical dash lines
    indicate the predicted epochs of periastron passage of the pulsar 
    \citep{Kuhnel2013}.  The inset zooms up the profile of
    the outburst in 2012 November, on which  Suzaku data taken on 2012 November
    20-22 are superposed with circles.}
  \label{fig:maxi_lc}
\end{figure}

MAXI \citep{matsuoka_pasj2009} Gas Slit Camear (GSC;
\cite{Mihara2011}) has been monitoring the X-ray flux of GRO
J1008$-$57 since the operation started on 2009 August 15
(MJD=55058) \citep{Sugizaki2011}.  Figure \ref{fig:maxi_lc} shows the
obtained light curve until 2013 June (MJD$\sim$56450).  By 2012
September (MJD$\sim 56200$), five outbursts were detected periodically at
the same orbital phase close to the pulsar periastron passage.  Their
peak intensities in 4--10 keV are almost same at 0.1 Crab, which corresponds to a
0.5--100 keV luminosity of $L_{\rm X}\simeq 2\times 10^{37}$ erg s$^{-1}$ at 5.8 kpc
assuming the same spectral shape as in Suzaku observation (section 3.4).
Thus,
these outbursts are categorized into the normal-type ones \citep{Reig2011}.
On 2012 November 5, the source exhibited unexpected brightening at
an irregular orbital phase which is $\sim 0.3$ cycle after the periastron 
\citep{Nakajima2012}. The 4--10 keV intensity reached $\sim$ 0.45 Crab at
the maximum. Judging from the outburst phase and the peak luminosity,
it is categorized into the giant-type outburst \citep{Reig2011}.

%

\section{Suzaku Observation of 2012 Giant Outburst
and Data Analysis}

\subsection{Observation and Data Reduction}

Triggered by the MAXI detection of the giant outburst from GRO
J1008$-$57, we requested a Suzaku ToO (Target of Opportunity)
observation.  It was performed on 2012 November 20, nearly coincident with  the outburst
maximum.  Suzaku covers an energy band from 0.5 to 500 keV with the
X-ray Imaging Spectrometer (XIS: \cite{Koyama2007}) and the Hard X-ray
Detector (HXD: \cite{Takahashi2007}, \cite{Kokubun2007}).  The target
was placed at the XIS nominal position on the focal-plane.
The XIS was operated in the normal mode with 1/4-window and 0.3 s
burst options, which affords a time resolution of 2 s.  The HXD was
operated in the nominal mode.  Table \ref{tab:table1} summarizes
the Suzaku observations including exposure and count rate in each
instrument.

The data reduction and analysis were performed with the standard
procedure using the Suzaku analysis software in HEASOFT version 6.12
and the CALDB files version 20110913, provided by the NASA/GSFC Suzaku
GOF.  All obtained data were first reprocessed by a Suzaku software
tool, {\tt aepipeline} to utilize the latest calibration.  The net
exposures after the standard event-screening process were 18.1 ks with
the XIS and 50.4 ks with the HXD. Due to the 0.3 s burst option, the XIS
exposure is about one third of that of the HXD.

We started the XIS data analysis with the standard cleaned event files.
On-source event data were collected from a circular region of $240''$ radius around the source position
on the XIS CCD images, and background data from an annulus with the inner
and outer radii of  $300''$ and $420''$, respectively.
The pileup effect on each image pixel was estimated by the Suzaku
PileupTools\footnote{http://www-utheal.phys.s.u-tokyo.ac.jp/$^\sim$yamada/soft/ XISPileupDoc\_20120221/XIS\_PileupDoc\_20120220.html}.
We excluded pixels on the image core in which the estimated pileup fraction is
larger than 1\% \citep{Yamada2012}.

In the HXD data analysis, we created the background spectra with the
standard procedure, using the archived background files provided by
the Suzaku GOF.  The obtained HXD-PIN background includes 
contribution from the  Cosmic X-ray Background (CXB), while it is
negligible in the HXD-GSO data \citep{Fukazawa2009}.  After
subtracting the backgrounds, the source count rates became 
$14.61\pm 0.01$ counts s$^{-1}$ in the PIN 20--60 keV band, and
$0.90\pm 0.02$ counts s$^{-1}$ in the GSO 60--115 keV band.

\begin{table*}
\begin{center}
  \caption{Log of Suzaku Observation of GRO J1008$-$57 in the 2012 November Giant Outburst}
  \label{tab:table1}
  \begin{tabular}{ccccp{1pt}ccp{1pt}cc}
    \hline              
    \hline              
    Date        & Obs Time       & \multicolumn{2}{c}{XIS-FI (0.8--10 keV)}  & & \multicolumn{2}{c}{HXD-PIN (20--60 keV)} & & \multicolumn{2}{c}{HXD-GSO (60--115 keV)} \\
    \cline{3-4}\cline{6-7}\cline{9-10}
    (2012       & Start/End      & Exp.          & Rate                      & & Exp.           & Rate                    & & Exp.           & Rate                   \\
     Nov.)      &   (UT)         & (ks)          & (counts s$^{-1}$)         & & (ks)           & (counts s$^{-1}$)       & & (ks)           & (counts s$^{-1}$)      \\
    \hline
    20--22      & 14:44/05:21    & 18.09         & 106.1$\pm$0.1             & & 50.38          & 14.61$\pm$0.01          & & 50.38          & 0.90$\pm$0.02         \\
    \hline
  \end{tabular}
\end{center}
\hspace*{1cm} Observation ID $=$ 907006010
\end{table*}


\subsection{Timing Analysis}
\label{sec:timinig}

\begin{figure}
  \begin{center}
    \includegraphics[width=80mm]{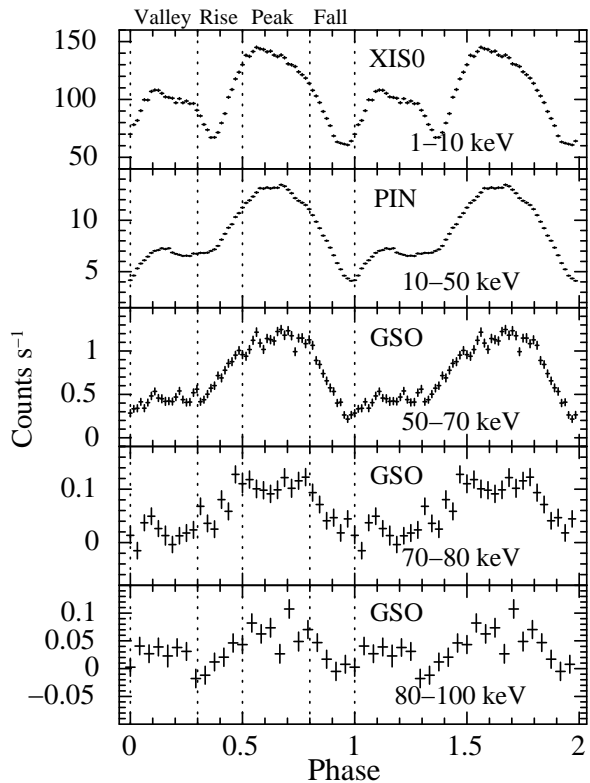}
  \end{center}
  \caption{ Folded pulse profiles by the XIS in 1--10 keV, 
    by HXD-PIN in 10--50 keV, and by HXD-GSO in the 50--70 keV, 70--80 keV, and 
    80--100 keV bands. 
    The dashed lines divide the pulse cycle into 
    Valley, Rise, Peak, and Fall, referring to the 50-70 keV profile.
    }
 \label{fig:pp}
\end{figure}

With the Suzaku analysis tool, {\tt aebarycen}, 
we converted the photon arrival times of all events into those at the
solar-system barycenter and
then searched the data for the coherent pulsation by epoch-folding analysis.
%
%
The $\sim$ 93.5 s pulsation was detected significantly, both with the XIS and the HXD, and 
the best period was obtained as 93.6257 $\pm$ 0.0001 s with
the HXD-PIN data.  
Figure \ref{fig:pp} shows the folded pulse
profiles in the XIS, HXD-PIN and HXD-GSO energy bands, where
the phase $\phi=0$ is set at the minimum in the HXD-PIN profile.
We divided the HXD-GSO band into three, 50--70 keV, 70--80 keV
and 80--100 keV, around the CRSF energy (section
\ref{sec:phase_average}).  The 50-70 keV profile is the same as that
of HXD-PIN, while that in 70--80 keV is somewhat different.
  The pulsation is still
significant in the highest energy band of 80-100 keV.

The pulse profile in the XIS has two peaks at $\phi\sim 0.1$ and
$\phi\sim 0.6$.  The former tends to decrease towards 
higher energies.  These double-peak profiles and their
energy dependence are largely consistent with the results obtained in
previous outbursts
\citep{Shrader1999,Coe2007,Naik2011,Kuhnel2013}.  
However, details
are rather different.  Comparing the XIS-band profiles, the former
peak obtained here is apparently smaller than that in the 2007
outburst.

%
%

As illustrated in figure \ref{fig:pp}, we divided the pulse cycle into four phases,
and named them Valley, Rise,
Peak, and Fall,  according to
the profile in the GSO 50--70 keV band.  They are used in the phase
resolved spectral analysis in section \ref{sec:phase_resolved}.


\subsection{Cyclotron Resonance Feature in Averaged Spectrum}
\label{sec:phase_average}


We examined a pulse-phase-averaged spectrum with the best photon statistics
for the previously suggested CRSF signatures.
All the spectral fitting attempts hereafter
were carried out on Xspec version 12.7.0.  The cross normalization
factor between the XIS and the HXD was fixed at 1:1.16
according to the latest calibration
information\footnote{http://www.astro.isas.ac.jp/suzaku/doc/suzakumemo/suzakumemo-2008-06.pdf}.
We discard the energy bands of 1.7--1.9 keV around the silicon K edge
and 2.1--2.4 keV around the gold M edges in the XIS data, where the
calibration uncertainty is relatively larger. 
We did not use the XIS-BI data either in the spectral analysis,
because it has larger calibration uncertainties than XIS-FI.

\begin{figure}
  \begin{center}
    \includegraphics[width=80mm]{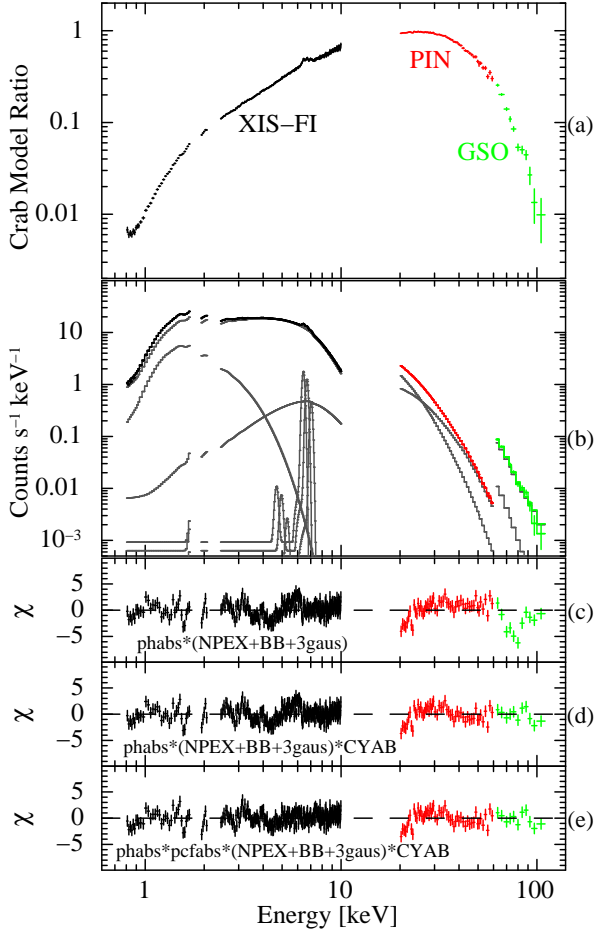}
  \end{center}
  \caption{
    Pulse-phase-averaged and background-subtracted broadband X-ray spectrum of GRO J1008$-$57
    by Suzaku XIS-FI (0.8-10 keV), HXD-PIN (20--60 keV) and HXD-GSO (60--115 keV).
    (a) Ratio to the Crab Nebula spectrum.
    (b) Count-rate spectra and the folded best-fit model of ${\tt pcfabs*phabs*(NPEX+BB+3gaus)*CYAB}$.
    (c) Residuals against ${\tt phabs*(NPEX+BB+3gaus)}$ model.
    (d) Residuals against ${\tt phabs*(NPEX+BB+3gaus)*CYAB}$ model.
    (e) Residuals against ${\tt pcfabs*phabs*(NPEX+BB+3gaus)*CYAB}$ model.
  }
 \label{fig:average_spec}
\end{figure}

Figure \ref{fig:average_spec} (a) shows ratios of the spectra obtained
with XIS-FI (0.8--10 keV), HXD-PIN (20--60 keV) and HXD-GSO
(60--115 keV) to those of the Crab nebula which has a simple
power-law shape with a photon index of $\sim 2.1$.  Figure
\ref{fig:average_spec} (b) shows the count-rate spectra
without removing instrument responses.  From the Crab ratios,
the spectrum is found to be largely approximated by a
smooth continuum with cutoffs below $\sim 2$ keV and above $\sim 20$
keV. In addition, iron-K emission lines at around 6.5 keV 
and an edge-like feature at around 70--80 keV are
clearly seen.
The latter looks like a typical CRSF observed in
some XBP spectra, and its energy is close to those of the possible absorption
features ($\sim 88$ keV) reported in past outbursts
\citep{Shrader1999,Kuhnel2013}.

\begin{figure}
  \begin{center}
    \includegraphics[width=85mm]{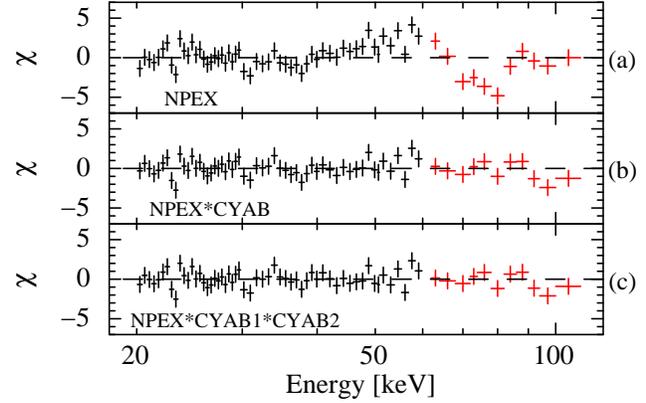}
  \end{center}
  \caption{
    Data-to-model residuals when the HXD data are 
    fitted with (a) NPEX alone, (b) NPEX*CYAB, and (c) NPEX*CYAB1*CYAB2
    in 20--115 keV band.
  }
 \label{fig:cyab_residual}
\end{figure}

We fitted the spectrum above 20 keV with typical XBP continuum models;
cutoff power-law (CPL, ${\tt cutoffpl}$ in Xspec terminology), FDCO
(Fermi-Dirac cutoff power-law; \cite{Tanaka1986}), and NPEX (Negative
and Positive power laws with EXponential cutoff: \cite{Mihara1998})
whose positive power-law index was fixed at 2.0.  
However, as exemplified in figure \ref{fig:cyab_residual} (a),
none of these models
alone were able to fit the data sufficiently, because of the feature at 70--80 keV.
We thus applied a cyclotron absorption factor (CYAB; $\tt cyclabs$ in Xspec
terminology, \cite{Mihara1990}, \cite{Makishima1999}) to the above continuum models.  
Since the width $W$ of the CYAB factor
cannot be constrained lower than the energy
resolution, $\sim 5$ keV at 80 keV in HXD GSO, we set its lower
limit at 2 keV in the model fits.
Then, all three continuum models became
acceptable within 90\% confidence limits, and the improvements of
chi-squared ($\chi^2_\nu$) for degree of freedom
($\nu$) were estimated 
with the F-test
to be significant above the 99\%
confidence limit .  
The case with NPEX * CYAB is shown in figure \ref{fig:cyab_residual} (b).
As listed in table \ref{tab:cyab},
the best-fit CRSF energy, $E_{\rm a} \sim $ 75--80 keV, slightly depends on the continuum model.

Since the CRSF energy, $\sim$ 80 keV, is rather high,
we should examine the possibility that it is in reality the second harmonic.
Actually, Vela X-1 has  been sometimes reported to show
a shallow absorption feature at $\sim 25$ keV \citep{Makishima1999},
possibly interpreted as the fundamental CRSF,
in addition to the more prominent feature at $\sim 50$ keV
which is confirmed in many observations 
(\cite{Mihara1998,Orlandini1998,Makishima1999,Kreykenbohm1999,Kreykenbohm2002,Odaka2013}).
We hence fitted the Suzaku spectra of GRO J1008$-$57
by a pair of {\it harmonic} CYAB factors,
with the fundamental resonance energy around $\sim$ 40 keV.
Then, as shown in figure 4(c) and given in table \ref{tab:cyab},
the best-fit $\chi^2_\nu$  slightly decreased  to 1.08
from 1.14 of the initial single-CYAB model,
yielding $E_1 \sim 37$ keV.
However, we consider this harmonic interpretation
rather unlikely for the following reasons.
First, such a local feature at $\sim 40$ keV is not visible in figure 4(b).
Second, an $F$-test indicates that the fit improvement
by introducing the second CRSF factor is less significant than 80\%.
Third, the derived ratio $D_1/D_2 =0.02$ of GRO J1008$-$57 is even smaller
than that of Vela X-1, $D_1/D_2=0.07/0.8=0.09$  \citep{Makishima1999}.
Finally, the obtained width $W_1 \sim 11$ keV for the lower-energy feature
is much wider than those of Vela X-1 ($W_1 \sim 2.2$ keV)
and the higher-energy feature of GRO J1008$-$57 ($W_2 = 2.0$ keV).
Therefore, we consider that the deep 75--80 keV feature GRO J1008$-$57
is the fundamental resonance,
although the alternate interpretation,
that it is the second harmonic resonance,
cannot be completely ruled out.

%
%



\begin{table*}
  \small
  \begin{center}
    \caption{Summary of model fits to Suzaku HXD 20--115 keV spectrum with either CPL, FDCO, or NPEX continuum and zero, one, or two CYABs.}
    \label{tab:cyab}
    \begin{tabular}{l@{\hspace{1mm}}ccccccccc}
      \hline
      \hline
      Continuum                & \multicolumn{3}{c}{CPL}                                         & \multicolumn{3}{c}{FDCO}                                       & \multicolumn{3}{c}{NPEX} \\
                               & None      & 1 CYAB                   & 2 CYAB                   & None     & 1 CYAB                   & 2 CYAB                   & None     & 1 CYAB                   & 2 CYAB \\
      \hline
      $\alpha_1$               & $ -0.63 $ & $ 0.11_{-0.13}^{+0.16} $ & $ 0.49_{-0.27}^{+0.64} $ & $ 1.61 $ & $ 0.69_{-0.12}^{+0.67} $ & $ 0.89_{-0.35}^{+0.43} $ & $ 1.25 $ & $ 0.16_{-0.22}^{+0.22} $ & $ 0.76_{-0.66}^{+0.71} $ \\
      $E_{\rm cut}$ (keV)       & $ 9.3   $ & $ 12.6_{-0.8}^{+1.2}   $ & $ 16.5_{-4.2}^{+3.8}   $ & $ 39.0 $ & $ 0.3_{-0.3}^{+28.9}   $ & $ 0.01_{-0.01}^{+155}  $ & ---      & ---                      & ---                      \\
      $E_{\rm fold}/kT$ (keV)   & ---       & ---                      & ---                      & $ 10.9 $ & $ 13.8_{-0.8}^{+0.3}   $ & $ 18.3_{-4.1}^{+5.3}   $ & $ 7.38 $ & $ 8.15_{-0.21}^{+0.671} $ & $ 7.81_{-0.18}^{+0.52} $ \\
      $A_1^{*}$                    & $ 0.03  $ & $ 0.15_{-0.04}^{+0.06} $ & $ 0.32_{-0.10}^{+0.53} $ & $ 3.11 $ & $ 0.57_{-0.14}^{+0.21} $ & $ 1.25_{-0.40}^{+1.93} $ & $ 4.73 $ & $ 0.24_{-0.11}^{+0.20} $ & $ 1.19_{-0.98}^{+2.41} $ \\
      $A_2^{*}$                    & ---       & ---                      & ---                      & ---      & ---                      & ---                      & $ 5.2  $ & $ 2.5_{-1.2}^{+0.6}    $ & $ 3.9_{-1.6}^{+0.9}    $ \\
      \hline
      $D_1$                    & ---       & $ 1.44_{-0.18}^{+0.20} $ & $ 0.06_{-0.03}^{+0.12} $ & ---      & $ 1.50_{-0.18}^{+0.21} $ & $ 0.08_{-0.04}^{+0.14} $ & ---      & $ 2.96_{-1.94}^{-1.06}$ & $ 0.06_{-0.03}^{+0.08} $ \\
      $E_{\rm a1}$ (keV)        & ---       & $ 79.5_{-2.2}^{+2.9}   $ & $ 40.7_{-1.3}^{+1.1}   $ & ---      & $ 80.0_{-2.0}^{+2.6}   $ & $ 40.6_{-1.0}^{+1.1}   $ & ---      & $ 74.4_{-1.3}^{+2.5}   $ & $ 36.8_{-0.7}^{+1.1}   $ \\
      $W_{1}$ (keV)            & ---       & $ 13.4_{-4.9}^{+6.7}   $ & $ 9.0_{-9.0}^{+11.4}   $ & ---      & $ 14.1_{-4.0}^{+4.0}   $ & $ 10.1_{-8.5}^{+10.0}  $ & ---      & $ 2.0_{-***}^{+6.4}    $ & $ 11.1_{-10.2}^{+7.2}   $ \\
      $D_2$                    & ---       & ---                      & $ 2.01_{-0.58}^{+1.13} $ & ---      & ---                      & $ 2.08_{-0.57}^{+1.16} $ & ---      & ---                      & $ 2.63_{-1.3}^{-1.0}   $ \\
      $E_{\rm a2}=2E_{\rm a1}$   & ---       & ---                      & $ 81.4                 $ & ---      & ---                      & $ 81.2                 $ & ---      & ---                      & $ 73.6                 $ \\
      $W_{2}$ (keV)            & ---       & ---                      & $ 23.7_{-7.2}^{+9.5}   $ & ---      & ---                      & $ 24.2_{-7.9}^{+11.2}  $ & ---      & ---                      & $ 2.0_{-***}^{+7.7}   $ \\
      \hline
      $\chi^2_{\nu}$ $(\nu)$   & $ 6.12 (64) $ & $ 1.25 (61) $ & $ 1.24 (59) $ & $ 2.59 (63) $ & $ 1.29 (60) $ & $ 1.02 (58) $ & $ 2.73 (63) $ & $ 1.14 (60) $ & $ 1.08 (58) $                            \\
      \hline
    \end{tabular}
    
  \end{center}
  
  \begin{tabular}{l@{\hspace{1mm}}l}
    \small
  $^{*}$ & Units in photons s$^{-1}$ cm$^{-2}$ keV$^{-1}$ at 1 keV.
  \end{tabular}
    
\end{table*}

\subsection{Broadband Spectral Model for Averaged Spectrum}

Now that the 20-100 keV HXD spectrum was successfully modeled and the CRSF was clearly
detected, the next step is to search for
broadband emission models that can explain the
whole Suzaku spectrum from 0.5 keV to 115 keV.
We first tested the CPL, FDCO, and NPEX continuum models as used in the
previous section, incorporating a CYAB factor at 75--80 keV and an
interstellar absorption (${\tt phabs}$ in Xspec terminology) whose
hydrogen column density $N_{\rm H}$ was set free.
In any continuum model, however, the fit was far from acceptable, and the
data-to-model residuals showed an excess in the soft X-ray band below
3 keV and iron K-lines at around 6.5 keV.  This
agrees with the results obtained in past outbursts
(e.g. \cite{Naik2011,Kuhnel2013}).

We then added a blackbody ($\tt BB$) model to account for the soft X-ray residuals,
and three narrow gaussians (${\tt gaus}$) for K$_\alpha$ lines
from neutral iron (6.4 keV) and helium-like iron (6.7 keV), as well as K$_\beta$
line at 7.05 keV.
%
%
Among the three continuum models, the NPEX-based composite model,
expressed by $\tt phabs*(NPEX+BB+3gaus)*CYAB$, fit the data much better
than the other two, but it is still unacceptable with $\chi^{2}_{\nu}
= 1.80$ for $\nu = 350$ degree of freedom.  The residuals, as shown in
figure \ref{fig:average_spec} (d), indicate that discrepancy remains
at around 4 keV and around 20 keV. To improve the fit, we tried to apply a partially
covering absorption model ($\tt pcfabs$ in Xspec terminology).  The
fit became even better with $\chi^{2}_{\nu} = 1.45$ for $\nu = 348$ as
shown in figure \ref{fig:average_spec} (e).
However, it is still outside the 90\% confidence limit.  This may
be because the phase-averaged spectrum has complex features that can arise by averaging pulse-phase dependent spectra.
%
%
Table \ref{tab:table2} summarizes all the best-fit model parameters.



\begin{table*}
 \small
  \begin{center}
    \caption{Best-fit models for Suzaku 0.8--115 keV broadband spectrum}
    \label{tab:table2}
    \begin{tabular}{cl@{\hspace{5mm}}c@{\hspace{5mm}}c@{\hspace{5mm}}c@{\hspace{5mm}}c}
      \hline
      \hline
               &          & \multicolumn{4}{c}{Model Function}  \\ 
               &                                                       & \cleft{${\tt phabs}$}           & \cleft{${\tt phabs}$}       & \cleft{${\tt phabs}$}       & \cleft{${\tt phabs*pcfabs}$} \\ 
      Component& Parameter                                             & \cleft{${\tt *Cont(CPL)}^\ddagger$}      & \cleft{${\tt *Cont(FDCO)}^\ddagger$} & \cleft{${\tt *Cont(NPEX)}^\ddagger$} & \cleft{${\tt *Cont(NPEX)}^\ddagger$} \\ 
               &                                                       & \cleft{${\tt *CYAB}$}           & \cleft{${\tt *CYAB}$}       & \cleft{${\tt *CYAB}$}       & \cleft{${\tt *CYAB}$}        \\ 
      \hline					     	    					      				      		       		   	   					         
      ${\rm phabs}$     & $N_{\rm H}$ ($10^{22}$ cm$^{-2}$)                 & $ 1.01 $            & $ 1.07 $             & $ 0.95_{-0.01}^{+0.02} $  & $ 0.93_{-0.02}^{+0.02} $           \\ 
      ${\rm bbody}$     & $kT_{\rm BB}$ (keV)                              & $ 0.36 $            & $ 0.30 $             & $ 0.41_{-0.01}^{+0.01} $  & $ 0.44_{-0.02}^{+0.02} $           \\ 
                        & $I_{\rm BB}^*$ ($\times 10^{-3}$)                 & $ 5.1  $            & $ 4.8  $              & $ 6.1_{-0.3}^{+0.3}    $ & $ 4.8_{-0.4}^{+0.4}    $           \\ 
      ${\rm gaus1}$     & $E_{\rm Fe \ K_{\alpha}}$ (keV)                    & $ 6.40 $            & $ 6.40 $             & $ 6.41_{-0.01}^{+0.01} $  & $ 6.42_{-0.01}^{+0.01} $           \\
                        & $I_{\rm Fe \ K_{\alpha}}^*$ ($\times 10^{-3}$)      & $ 3.9  $            & $ 4.2  $              & $ 3.7_{-0.2}^{+0.2}    $ & $ 3.0_{-0.2}^{+0.3}    $           \\
      ${\rm gaus2}$     & $E_{\rm Fe \ 6.7}$ (keV)                          & $ 6.69 $            & $ 6.69 $             & $ 6.69_{-0.01}^{+0.01} $   & $ 6.69_{-0.01}^{+0.02} $           \\
                        & $I_{\rm Fe \ 6.7}^*$ ($\times 10^{-3}$)            & $ 3.3  $            & $ 3.7  $              & $ 3.0_{-0.2}^{+0.2}    $  & $ 2.3_{-0.3}^{+0.2}    $           \\
      ${\rm gaus3}$     & $E_{\rm Fe \ K_{\beta}}$ (keV)                     & $ 7.00 $            & $ 7.01 $             & $ 7.00_{-0.02}^{+0.02} $  & $ 7.04_{-0.03}^{+0.04} $           \\
                        & $I_{\rm Fe \ K_{\beta}}^*$ ($\times 10^{-4}$)       & $ 17.4 $            & $ 21.1 $              & $ 14.6_{-2.2}^{+2.1}   $  & $ 9.5_{-2.2}^{+2.2}    $           \\
      ${\rm pcfabs}$    & $N_{\rm H}$ ($10^{22}$ cm$^{-2}$)                  & ---                 & ---                   & ---                      & $ 32.5_{-4.8}^{+4.4}   $           \\ 
                        & $f_{\rm PCF}$                                     & ---                 & ---                   & ---                      & $ 0.18_{-0.02}^{+0.02} $          \\ 
      ${\rm NPEX}$      & $\alpha_1$                                       & $ 0.51 $            & $ 0.69 $              & $ 0.27_{-0.01}^{+0.01} $ & $ 0.38_{-0.02}^{+0.03} $            \\ 
                        & $E_{\rm cut}$ (keV)                               & $ 14.7 $            & $ 0.0  $            & ---			   & ---                               \\
                        & $kT/E_{\rm fold}$ (keV)                           & ---                 & $ 13.8 $            & $ 8.79_{-0.23}^{+0.34} $ & $ 8.58_{-0.25}^{+0.60} $            \\ 
                        & $A_{1}$$^{\dagger}$ ($\times 10^{0}$)              & $ 0.34 $            & $ 0.79 $            & $ 0.29_{-0.00}^{+0.00} $ & $ 0.39_{-0.02}^{+0.02} $            \\
                        & $A_{2}$$^{\dagger}$ ($\times 10^{-4}$)             & ---                 & ---                 & $ 1.4_{-0.2}^{+0.2}    $ & $ 1.6_{-0.4}^{+0.3}    $            \\ 
      ${\rm CYAB}$      & $D$                                              & $ 1.55 $            & $ 1.58 $             & $ 0.88_{-0.15}^{+0.39} $ & $ 0.79_{-0.14}^{+1.26} $              \\
                        & $E_{\rm a}$ (keV)                                 & $ 80.2 $            & $ 79.8 $            & $ 78.1_{-2.8}^{+4.1}   $ & $ 78.1_{-3.6}^{+7.6}   $              \\
                        & $W$ (keV)                                        & $ 11.9 $            & $ 6.2 $             & $ 11.6_{-6.9}^{+8.7} $   & $ 11.8_{-9.6}^{+19.7} $                     \\
\hline
                        & $\chi^2_{\nu}$ $(\nu)$                            & 2.53 (351)         &  3.91 (350)       &  1.80 (350)            &  1.45 (348)                      \\
                        & $L_{\rm 0.5-100 \ keV}$$^\S$                      & ---                  & ---                 &    ---                 & $ 10.93_{-0.05}^{+0.01} $  \\
																      
\hline
    \end{tabular}

  \end{center}

  \begin{tabular}{l@{\hspace{1mm}}l}
  $^{*}$ & Units in photons s$^{-1}$ cm$^{-2}$.\\
  $^{\dagger}$ & Units in photons s$^{-1}$ cm$^{-2}$ keV$^{-1}$ at 1 keV.\\
  $^{\ddagger}$ & ${\tt Cont(CPL)}$ is a CPL-based composite model including a soft {\tt BB} and three iron-lines, expressed by ${\tt CPL+BB+3gaus}$.\\
  & ${\tt Cont(FDCO)}$ and ${\tt Cont(NEPX)}$ represent FDCO-based and NPEX-based composite models, respectivly.\\
  $^\S$ & Units in $ 10^{37}$ erg s$^{-1}$.

  \end{tabular}
    
\end{table*}

\subsection{Pulse-Phase Resolved Spectra}
\label{sec:phase_resolved}

\begin{figure}
  \begin{center}
    \includegraphics[width=80mm]{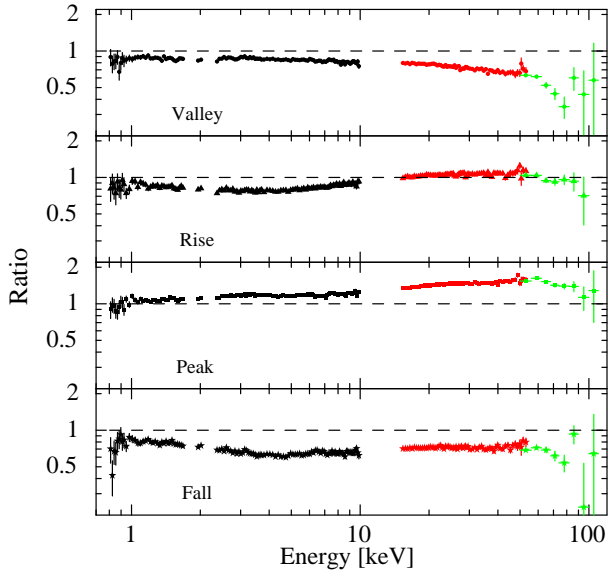}
  \end{center}
  \caption{ Pulse-phase resolved spectra for Valley, Rise, Peak, and Fall
    phases in
    the form of the ratio to the model that best describes the pulse-phase averaged spectrum.
 }
 \label{fig:resolved_ratio}
\end{figure}

As seen in figure \ref{fig:pp}, the folded pulse profiles 
from 1 keV to 100 keV are apparently energy dependent.  This
means that the energy spectrum depends on the pulse phase.
We thus extracted four spectra, one from each of the four pulse phases defined in figure
\ref{fig:pp} from the 50--70 keV GSO pulse profile.
Figure \ref{fig:resolved_ratio}
shows the obtained four
spectra in a form of their ratios to the phase-averaged spectrum.  While
the spectrum in the Valley phase is softer than the average, that in
the Peak phase is harder.
No feature are apparent at the iron K-line band around 6--7 keV
in these ratio plots.
This indicates that the equivalent width of the iron lines does not
depend on the pulse phases.

We fitted each phase-resolved spectrum with the model which best
described the phase averaged spectrum in section
\ref{sec:phase_average}. Here, we fixed the iron K$_\beta$-line energy at
7.05 keV and the CYAB width at a typical value of 5.0 keV
because they were poorly constrained by the data with lower statistics.
The fits became acceptable except for the Peak phase.
Table \ref{tab:resolved_param} summarizes the obtained best-fit
parameters in each phase.

\begin{figure}
  \begin{center}
    \includegraphics[width=70mm]{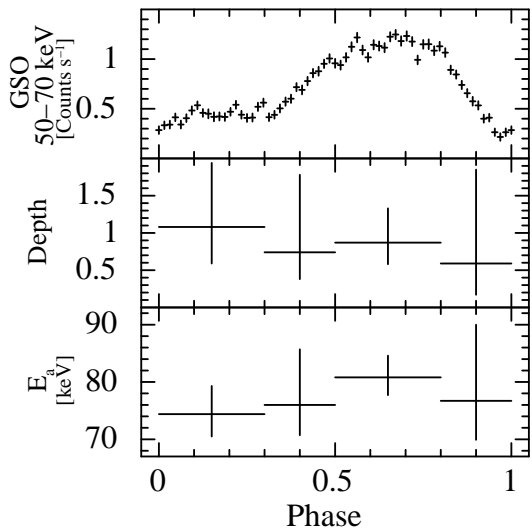}
  \end{center}
  \caption{ Pulse-phase dependence of the CRSF depth $D$ and energy
    $E_{\rm a}$.  The vertical error bars of
    $D$ and $E_{\rm a}$ represent 90\% confidence limits of statitical
    uncertainties.  }
 \label{fig:resolved_spec}
\end{figure}

Although these model parameters for the continuum are
correlated in complex ways, those of the CRSF model are mostly free from them.  
As shown in figure \ref{fig:resolved_spec}, the derived CRSF parameters 
show some dependence on the pulse phase,
but not more significantly than errors.


\begin{table*}
  \small
  \begin{center}
    \caption{Best fit parameters of phase resolved spectra}
    \label{tab:resolved_param}
    \begin{tabular}{l@{\hspace{1cm}}l@{\hspace{1cm}}c@{\hspace{1cm}}c@{\hspace{1cm}}c@{\hspace{1cm}}c}
      \hline
      \hline
                       &                                      & \multicolumn{4}{c}{Pulse Phase}                                                                           \\
      Component        & Parameter                                 & Valley                   & Rise                     & Peak                     & Fall                     \\
      \hline					       		  	  			  	     		  	     			       
      ${\rm phabs}$   &   $N_{\rm H1}$ ($10^{22}$ cm$^{-2}$)           & $ 1.01_{-0.06}^{+0.06} $ & $ 0.90_{-0.05}^{+0.05} $ & $ 1.01_{-0.04}^{+0.04} $ & $ 0.92_{-0.05}^{+0.05} $  \\
      ${\rm bbody}$    &  $kT_{\rm BB}$ (keV)                         & $ 0.30_{-0.03}^{+0.05} $ & $ 0.43_{-0.03}^{+0.04} $ & $ 0.36_{-0.03}^{+0.04} $ & $ 0.43_{-0.03}^{+0.04} $  \\
                       &   $I_{\rm BB}$ ($\times 10^{-3}$)              & $ 3.3_{-1.0}^{+1.4}    $ & $ 5.5_{-0.7}^{+0.7}    $ & $ 4.1_{-0.8}^{+0.9}    $ & $ 5.5_{-0.7}^{+0.7}    $  \\
      ${\rm gaus_{1}}$ & $E_{\rm Fe \ K_{\alpha}}$ (keV)                & $ 6.40_{-0.03}^{+0.04} $ & $ 6.41_{-0.03}^{+0.03} $ & $ 6.41_{-0.03}^{+0.03} $ & $ 6.43_{-0.02}^{+0.02} $  \\
                       & $I_{\rm Fe \ K_{\alpha}}$$^{*}$ ($\times 10^{-3}$)   & $ 2.0_{-0.4}^{+0.5}    $ & $ 2.6_{-0.5}^{+0.6}    $ & $ 2.6_{-0.5}^{+0.5}    $ & $ 3.3_{-0.5}^{+0.5}    $  \\
      ${\rm gaus_{2}}$ & $E_{\rm Fe \ 6.7}$ (keV)                       & $ 6.66_{-0.03}^{+0.04} $ & $ 6.69_{-0.04}^{+0.04} $ & $ 6.68_{-0.03}^{+0.04} $ & $ 6.74_{-0.05}^{+0.05} $  \\
                       & $I_{\rm Fe \ 6.7}$$^{*}$ ($\times 10^{-3}$)          & $ 1.9_{-0.5}^{+0.4}    $ & $ 2.2_{-0.6}^{+0.6}    $ & $ 2.0_{-0.3}^{+0.5}    $ & $ 1.7_{-0.5}^{+0.5}    $  \\
      ${\rm gaus_{3}}$ & $I_{\rm Fe \ K_{\beta}}$$^{*}$ ($\times 10^{-4}$)    & $ 6.9_{-3.9}^{+3.8}    $ & $ 7.2_{-4.9}^{+4.8}    $ & $ 1.3_{-1.3}^{+4.5}    $ & $ 6.6_{-4.5}^{+4.5}    $  \\
      ${\rm pcfabs}$   & $N_{\rm H2}$ ($10^{22}$ cm$^{-2}$)             & $ 48.3_{-5.9}^{+6.2}   $ & $ 42.3_{-14.7}^{+12.7} $ & $ 53.8_{-6.2}^{+7.0}   $ & $ 41.6_{-10.3}^{+8.4}  $  \\
                       & $f_{\rm PCF}$                                  & $ 0.23_{-0.05}^{+0.04} $ & $ 0.18_{-0.07}^{+0.06} $ & $ 0.22_{-0.04}^{+0.04} $ & $ 0.28_{-0.06}^{+0.06} $  \\
      ${\rm NPEX}$     & $\alpha_1$                                     & $ 0.48_{-0.04}^{+0.04} $ & $ 0.24_{-0.05}^{+0.05} $ & $ 0.41_{-0.03}^{+0.03} $ & $ 0.40_{-0.06}^{+0.06} $  \\
                       &  $kT$              (keV)                        & $ 7.78_{-0.13}^{+0.15} $ & $ 8.08_{-0.13}^{+0.15} $ & $ 8.11_{-0.07}^{+0.07} $ & $ 8.03_{-0.14}^{+0.17} $  \\
                       & $A_{1}$$^{\dagger}$ ($\times 10^{0}$)                & $ 0.46_{-0.05}^{+0.05} $ & $ 0.26_{-0.04}^{+0.04} $ & $ 0.54_{-0.05}^{+0.05} $ & $ 0.30_{-0.05}^{+0.06} $  \\
                       & $A_{2}$$^{\dagger}$ ($\times 10^{-4}$)               & $ 2.0_{-0.2}^{+0.2}    $ & $ 2.5_{-0.3}^{+0.3}    $ & $ 3.8_{-0.2}^{+0.2}    $ & $ 1.8_{-0.2}^{+0.2}    $  \\
      ${\rm CYAB}$     & $D$                                            & $ 1.08_{-0.49}^{+0.86} $ & $ 0.74_{-0.36}^{+1.04} $ & $ 0.87_{-0.29}^{+0.46} $ & $ 0.59_{-0.42}^{+1.26} $  \\
                       & $E_{\rm a}$ (keV)                              & $ 74.4_{-3.9}^{+4.9}   $ & $ 76.0_{-5.3}^{+9.7}   $ & $ 80.8_{-3.1}^{+3.8}   $ & $ 76.7_{-6.8}^{+13.3}  $  \\
\hline
                       & $\chi^2_{\nu}$ $(\nu)$                         &  1.13 (171)            &  0.99 (171)            &  1.49 (171)            &  1.07 (171)             \\
                       & $L_{\rm 0.5-100 \ keV}$$^{\ddagger}$           & $ 8.86_{-0.09}^{+0.04} $ & $ 11.10_{-0.13}^{+0.04} $ & $ 15.66_{-0.09}^{+0.05} $ & $ 8.00_{-0.16}^{+0.04} $  \\
\hline
    \end{tabular}

  \end{center}

  \begin{tabular}{l@{\hspace{1mm}}l}
    \multicolumn{2}{l}{Spectal model function: ${\tt phabs*pcfabs*(NPEX+BB+3 gaus)*CYAB}$}                                             \\
    \multicolumn{2}{l}{Energy of iron K$_{\beta}$ is fixed to 7.05 keV. Width of CYAB is fixed to 5.0 keV.}\\
    $^{*}$ & Units in photons s$^{-1}$ cm$^{-2}$.\\
    $^{\dagger}$ & Units in photons s$^{-1}$ cm$^{-2}$ keV$^{-1}$ at 1 keV. \\
    $^{\ddagger}$ & Units in $ 10^{37}$ erg s$^{-1}$.
  \end{tabular}

  %

\end{table*}

\section{Discussion}

\subsection{Possible CRSF Energy Change}

We analyzed the broadband X-ray (0.8--115 keV) spectrum of GRO
J1008$-$57 obtained by Suzaku, covering the peak of the giant outburst
in 2012 detected by MAXI, and found a significant absorption
signature at 75--80 keV  \citep{Yamamoto2013}.  
It can be interpreted as a fundamental CRSF, and reconfirms, with much higher significance,
the previous suggestions (\cite{Shrader1999}, \cite{Kuhnel2013}).

\begin{figure}
  \begin{center}
    \includegraphics[width=70mm]{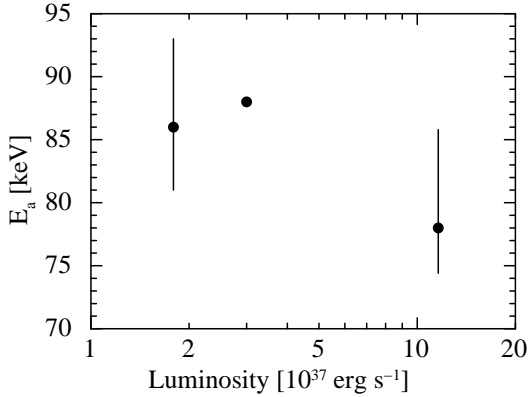}
  \end{center}
  \caption{Luminosity (0.5--100 keV) dependence of the CRSF energy 
    from table \ref{tab:lxea}.}
 \label{fig:lxea}
\end{figure}

Table \ref{tab:lxea} compares the CRSF parameters and
luminosity obtained in this work with those of the previous outbursts,
and figure \ref{fig:lxea} gives its graphical plot.
Thus, the CRSF energy might decrease towards higher luminosities,
although the presently available information is very limited.
%

The luminosity dependence of the CRSF energy has been observed
in several XBPs.  While some of them, 4U 0115$+$63 \citep{Mihara1998,
  Mihara2004, Nakajima2006} and V 0332$+$53 \citep{Tsygankov2006,
  Mowlavi2006, Nakajima2010}, showed negative correlations,
others, Her X-1 \citep{Gruber2001, Staubert2007} and GX 304$-$1
\citep{Yamamoto2011,Klochkov2012} showed positive.
These are explained by variations of the
cyclotron-scattering photosphere; it increases by radiation
pressure in the super-Eddington luminosity regime
\citep{Mihara1998}, while it decreases due to dynamical pressure of the
accretion in the sub-Eddington luminosity \citep{Staubert2007}.
The luminosity of GRO J1008$-$57 observed by Suzaku at the peak of the 2012 giant
outburst, $1.1\times 10^{38}$ erg s$^{-1}$, is close to the Eddington
luminosity for the typical neutron-star mass of $1.4\MO$.  Therefore,
the possible CRSF energy change suggests such a situation that the
accretion mode changed from the sub-Eddington to the super-Eddington
regime at that time.

%
%

%

\begin{table}
  \begin{center}
    \caption{CRSF measurements in GRO J1008$-$57.}
    \label{tab:lxea}
    \small
    \begin{tabular}{lcccc}
      \hline
      \hline

                  &                        & \multicolumn{3}{c}{CRSF parameters} \\
      Outburst    & Luminosity$^{*}$       &  $E_{a}$     & $W$         & $D$ \\
                  & (10$^{37}$ erg s$^{-1}$) &             &  (keV)      &       \\
      \hline                                                      
      2012 Nov.$^{\dagger}$  &  10.93                & $76^{+1.9}_{-1.7}$  & 5 (fix)  &  $1.08^{+0.25}_{-0.21}$ \\
      2007 Dec.$^{\ddagger}$ &  1.79                & $86^{+7}_{-5}$      & $8^{+6}_{-4}$  & 2.3 (fix)\\
      1993 Jul.$^{\S}$      &  3.0                 & $88$               & ---    & $2.3^{+0.6}_{-0.6}$ \\
      \hline
    \end{tabular}
  \end{center}
  $^{*}$ Calculated from the best spectral model in 0.5--100 keV.\\
  $^{\dagger}$ This work, $^{\ddagger}$ \citet{Kuhnel2013}, $^{\S}$ \citet{Shrader1999}, 
\end{table}

\subsection{Magnetic Fields in Binary Pulsars}

\begin{figure}
  \begin{center}
    \includegraphics[width=85mm]{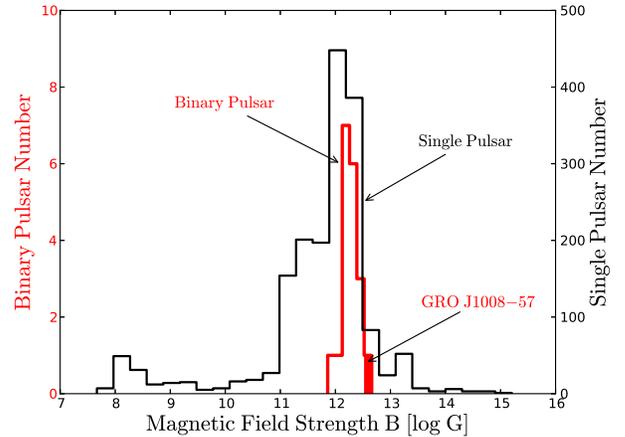}
  \end{center}
  \caption{ Distribution of magnetic field strengths on neutron stars
    in binary pulsars estimated from the CRSFs (red ordinate to the left) and 
    in single pulsars from the period and the period derivative from ATNF pulsar catalog
    \citep{2005AJ....129.1993M} (black, to the right).  
    It is updated from those in \citet{Mihara1998} and \citet{Makishima1999}.  
    }
 \label{fig:Bhist}
\end{figure}

%
We now know 18 XBPs in which CRSFs are significantly detected
and their parameters are determined well.  The CRSF energy of 75--80
keV obtained here from GRO J1008$-$57 is the highest among them.
Therefore, the estimated surface magnetic field, $6.6 \times 10^{12}$
$(1+z_g)$ G, extends the highest end of their magnetic field
distribution.
Figure \ref{fig:Bhist} shows the updated distribution of the XBP
magnetic field strengths.  
It is still clustered in a
very narrow range of $(1.0-6.6)\times 10^{12}$ G, compared to the
distribution of a larger number ($\sim 1000$) of single radio pulsars
in the ATNF pulsar catalog \citep{2005AJ....129.1993M}.  
%
%
Although the radio pulsars show considerably broader field distribution,
this could be due to the much lower accuracy of their field determinations 
which assume spin down via magnetic dipole radiation.
%
%
%
In any case, the plots favor the scenario 
that the surface magnetic fields of neither XBPs nor radio
pulsars would decay significantly within their lifetime of $\sim
10^8$ yr \citep{Itoh1995,Makishima1999}.
%
%

\begin{figure}
  \begin{center}
    \includegraphics[width=85mm]{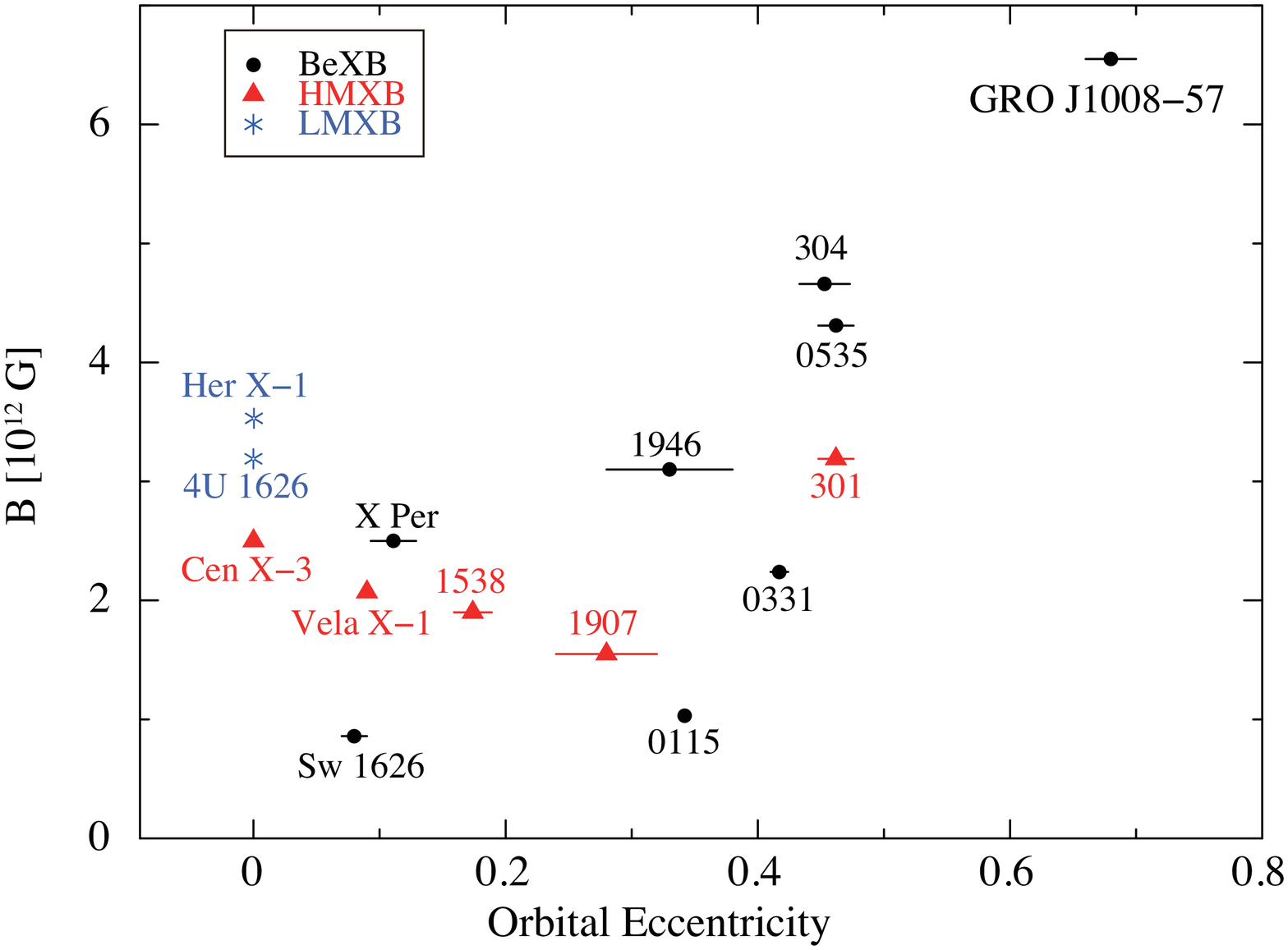}
  \end{center}
  \caption{ The magnetic field strengths of XBPs measured with the CRSF technique, plotted against
    their orbital eccentricities.
    Meanings and references of labels to the data points are as follows : 
    GRO J1008$-$57 (This work; \cite{Coe2007}),
    304 (GX 304$-$1; \cite{Yamamoto2011}; \cite{Yamamoto2013phD}; Yamamoto et al in prep.), 
    0535 (A 0535$+$26; \cite{Caballero2013}; \cite{Finger1994}),  
    Her X$-$1 (\cite{Vasco2011}; \cite{Staubert2009}), 
    4U 1626 (4U 1626$-$67; \cite{Iwakiri2012}; \cite{Chakrabarty1997}), 
    301  (GX 301$-$2; \cite{Suchy2012}; \cite{Koh1997}), 
    1946  (XTE J1946$+$274 ; \cite{Maitra2013}; \cite{Wilson2003}), 
    X Per (\cite{Coburn2001}; \cite{Delgado2001}), 
    Cen X$-$3 (\cite{Suchy2008}; \cite{Raichur2010a}), 
    0331  (X0331$+$53; \cite{Nakajima2010}; \cite{Raichur2010b}),
    Vela X$-$1 (\cite{Odaka2013}; \cite{Bildsten1997}),
    1538 (4U 1538$-$52; \cite{Rodes-Roca2009}; \cite{Clark2000}), 
    1907  (4U 1907$+$09; \cite{Rivers2010}; \cite{Baykal2006}),   
    0115  (4U 0115$+$63; \cite{Nakajima2006}; \cite{Raichur2010b}), 
    Sw 1626 (Swift J1626.6$-$5156; \cite{DeCesar2013}).
  }
 \label{fig:eccB}
\end{figure}

GRO J1008$-$57 is known to have a large orbital eccentricity of
$e=0.68(2)$ \citep{Coe2007}. In figure \ref{fig:eccB}, we plot a
relation between the surface magnetic fields and orbital
eccentricities of 15 XBPs whose CRSFs and binary orbital parameters
are well determined.  On this plot, GRO J1008$-$57 locates at the
upper right corner.
Thus, the surface magnetic field and the orbital
eccentricity of high mass X-ray binaries, including BeXBs,
appear to have a positive correlation.  This may suggest
the evolutional relation between these parameters.
%
%
%
%
%
Since the surface magmatic fields would not change as discussed above
and the orbital eccentricity would not change significantly within
their lifetime, the correlation is considered to be formed when the
XBPs are born.  Further observational as well as theoretical studies
are necessary.
%
%
%
%

%

\bigskip


We thank the Suzaku operation team for arranging and carrying out the
TOO observations.  We are also grateful to all members of the MAXI and
the ISS-operation teams.  
This research was partially supported by the
Ministry of Education, Culture, Sports, Science and Technology (MEXT),
Grant-in-Aid No. 24340041 and 23244024.



\end{document}